\documentstyle[]{mn}
\newif\ifAMStwofonts

\def\etal{{\rm et al.}}

\def\simgt{\mathrel{\spose{\lower 3pt\hbox{$\sim$}}
        \raise 2.0pt\hbox{$>$}}}
\def\simlt{\mathrel{\spose{\lower 3pt\hbox{$\sim$}}\raise 2.0pt\hbox{$<$}}}

\ifoldfss
  \newcommand{\rmn}[1] {{\rm #1}}

  \ifCUPmtlplainloaded \else
    \NewTextAlphabet{textbfit} {cmbxti10} {}
    \NewTextAlphabet{textbfss} {cmssbx10} {}
    \NewMathAlphabet{mathbfit} {cmbxti10} {} 
    \NewMathAlphabet{mathbfss} {cmssbx10} {} 
  \fi
  \ifAMStwofonts
    \ifCUPmtlplainloaded \else
      \NewSymbolFont{upmath} {eurm10}
      \NewSymbolFont{AMSa} {msam10}
      \NewMathSymbol{\upi}     {0}{upmath}{19}
      \NewMathSymbol{\umu}     {0}{upmath}{16}
      \NewMathSymbol{\upartial}{0}{upmath}{40}
      \NewMathSymbol{\leqslant}{3}{AMSa}{36}
      \NewMathSymbol{\geqslant}{3}{AMSa}{3E}

    \fi
  \fi
\fi 

\ifnfssone
  \newmathalphabet{\mathit}
  \addtoversion{normal}{\mathit}{cmr}{m}{it}
  \addtoversion{bold}{\mathit}{cmr}{bx}{it}
  \newcommand{\rmn}[1] {\mathrm{#1}}

  \newmathalphabet{\mathbfit} 
  \addtoversion{normal}{\mathbfit}{cmr}{bx}{it}
  \addtoversion{bold}{\mathbfit}{cmr}{bx}{it}
  \newmathalphabet{\mathbfss} 
  \addtoversion{normal}{\mathbfss}{cmss}{bx}{n}
  \addtoversion{bold}{\mathbfss}{cmss}{bx}{n}
  \ifAMStwofonts
    \ifCUPmtlplainloaded \else
      \UseAMStwoboldmath
      \makeatletter
      \new@mathgroup\upmath@group
      \define@mathgroup\mv@normal\upmath@group{eur}{m}{n}
      \define@mathgroup\mv@bold\upmath@group{eur}{b}{n}
      \edef\UPM{\hexnumber\upmath@group}
      \new@mathgroup\amsa@group
      \define@mathgroup\mv@normal\amsa@group{msa}{m}{n}
      \define@mathgroup\mv@bold\amsa@group{msa}{m}{n}
      \edef\AMSa{\hexnumber\amsa@group}
      \makeatother
      \mathchardef\upi="0\UPM19
      \mathchardef\umu="0\UPM16
      \mathchardef\upartial="0\UPM40
      \mathchardef\leqslant="3\AMSa36
      \mathchardef\geqslant="3\AMSa3E
    \fi
  \fi
\fi 

\ifnfsstwo
  \newcommand{\rmn}[1] {\mathrm{#1}}

  \DeclareMathAlphabet{\mathbfit}{OT1}{cmr}{bx}{it}
  \SetMathAlphabet\mathbfit{bold}{OT1}{cmr}{bx}{it}
  \DeclareMathAlphabet{\mathbfss}{OT1}{cmss}{bx}{n}
  \SetMathAlphabet\mathbfss{bold}{OT1}{cmss}{bx}{n}
  \ifAMStwofonts
    \ifCUPmtlplainloaded \else
      \DeclareSymbolFont{UPM}{U}{eur}{m}{n}
      \SetSymbolFont{UPM}{bold}{U}{eur}{b}{n}
      \DeclareSymbolFont{AMSa}{U}{msa}{m}{n}
      \DeclareMathSymbol{\upi}{0}{UPM}{"19}
      \DeclareMathSymbol{\umu}{0}{UPM}{"16}
      \DeclareMathSymbol{\upartial}{0}{UPM}{"40}
      \DeclareMathSymbol{\leqslant}{3}{AMSa}{"36}
      \DeclareMathSymbol{\geqslant}{3}{AMSa}{"3E}
    \fi
  \fi
\fi 

\ifCUPmtlplainloaded \else
  \ifAMStwofonts \else 
    \def\upi{\pi}
    \def\umu{\mu}
    \def\upartial{\partial}
  \fi
\fi

\title[Constraints on the mass-profile of the lens galaxy G2237+0305]
  {Constraints on the mass-profile of the lens galaxy G2237+0305}
\author[J. S. B. Wyithe et al.]
  {J.~S.~B.~Wyithe$^{1,2}$, 
  E.~Agol$^3$,
  E. L. Turner$^2$,
  R. W. Schmidt$^4$\\
  $^1$ School of Physics, University of Melbourne, Parkville, Vic, 3052, 
Australia\\
  $^2$ Princeton University Observatory, Peyton Hall, Princeton, NJ 08544, USA\\
  $^3$ Physics and Astronomy Department, Johns Hopkins University, Baltimore, MD 21218, USA\\
  $^4$ Institute for Astronomy, University of Cambridge, Madingly Road, Cambridge, CB3 0HA, UK\\
 Email: swyithe@astro.princeton.edu, agol@pha.jhu.edu, elt@astro.princeton.edu, rschmidt@ast.cam.ac.uk}
\date{Accepted Received}
\pagerange{\pageref{firstpage}--\pageref{lastpage}}
\pubyear{1999}

\def\LaTeX{L\kern-.36em\raise.3ex\hbox{a}\kern-.15em
    T\kern-.1667em\lower.7ex\hbox{E}\kern-.125emX}

\begin{document}

\label{firstpage}

\maketitle

\begin{abstract}

Published parametric models of the Einstein Cross gravitational lens demonstrate that the image geometry can be reproduced by families of models. In particular, the slope of the mass-profile for the lens galaxy is unconstrained. However, recent models predict a dependence of image flux ratios on the slope of the mass profile. We use this dependence to constrain the mass profile by calculating the likelihood of the slope using published mid-IR flux ratios (including microlensing variability). We find that the galaxy is likely to be flatter than isothermal, and therefore that the mass-to-light ratio is decreasing in the inner kpc.

\end{abstract}


\section{Introduction}

Q2237+0305 (The Einstein Cross) was discovered in the CfA Redshift survey (Huchra et al. 1985). The object comprises a quasar at redshift $z=1.695$ that is gravitationally lensed by a foreground galaxy ($z=0.0394$) producing 4 images with separations of $\sim 1''$. Many models have been proposed for the projected galaxy lens mass distribution based on observations of the lensed images (e.g. Kent \& Falco 1988, hereafter KF88; Schneider et al. 1988, hereafter S88; Kochanek 1991, hereafter K91; Rix, Schneider \& Bachall 1992, hereafter RSB92; Wambsganss \& Paczynski 1994, hereafter WP94; Witt, Mao \& Schechter 1995, hereafter WMS95; SWL98; Chae, Turnshek \& Khersonsky 1998, hereafter CTK98). Studies which employ parametric lens models having galaxy and image positions as constraints describe a degeneracy between the ellipticity of the mass profile and its slope $\nu$. As a result, the total magnification is virtually unconstrained. On the other hand, the models of SWL98 and CTK98 predict variation of the flux ratios with $\nu$ (this dependence arises due to the slightly asymmetric image geometry). While optical flux ratios cannot be used as model constraints due to microlensing and uncertain differential extinction (e.g. S88; KF88; K91; WP94), measurements in the radio (Falco et al. 1996) should be reliable due to an extended emission region. However Q2237+0305 is faint in the radio, and the observations have large uncertainties. As a result, the majority of published models predict flux ratios consistent with those measured in the radio. 

Agol, Jones \& Blaes (2000, hereafter AJB00) observed Q2237+0305 in the mid-IR and measured flux ratios with uncertainties significantly smaller than observations in the radio. Each of the 4 images are observed through the bulge of the lens galaxy which has an optical depth in stars that is of order unity (e.g. KF88; S88; SWL98). This results in a high probability for microlensing, and the mid-IR observations were coincident with significant microlensing of the optical flux. This disparity between optical and mid-IR flux ratios has been used to demonstrate that the mid-IR emission region is large with respect to the microlens Einstein Radius (AJB00; Wyithe, Agol \& Fluke 2001, hereafter WAF01). The large source size indicates that unlike the optical emission, the mid-IR emission is not subject to significant microlensing. In addition, extinction is not important in the mid-IR. The mid-IR flux ratios should therefore be reliable indicators of the true flux-ratios, and can be used to constrain galaxy mass models.

In Sec.~\ref{macros} we assume microlensing of the mid-IR is negligible and compute the likeli-hood for different published macro-models by comparing the observed mid-IR flux ratios with the predicted flux-ratios. In Sec.~\ref{MLlim} we discuss the effect of mid-IR microlensing on the  results obtained.

\section{Constraining Published Macro Models for Q2237+0305}
\label{macros}

\begin{table*}
\begin{center}
\caption{\label{tab1} Values for the image C flux at different epochs in the mid-IR (AJB00, Agol et al. 2001) and V-bands (Wozniak et al. 2000; OGLE web page).}
\begin{tabular}{|c|c|c|}
\hline
Wavelength         & Date           & Image C flux   \\\hline
11.7$\mu m$ & 28th July \& 21st September 1999 & 3.15 $\pm$.4 mJy\\
V-band             & 1st August 1999   & 0.46$\pm$.01 mJy\\ 
V-band             & 26th September 1999 & 0.34$\pm$.01 mJy\\\\

11.7$\mu m$ & 5th October  2000  & 3.55$\pm$.3 mJy\\ 
V-band      & 7th November 2000 & 0.180$\pm$.015 mJy\\\hline 

\end{tabular}
\end{center}

\end{table*}

\begin{figure*}
\vspace*{125mm}
\includegraphics{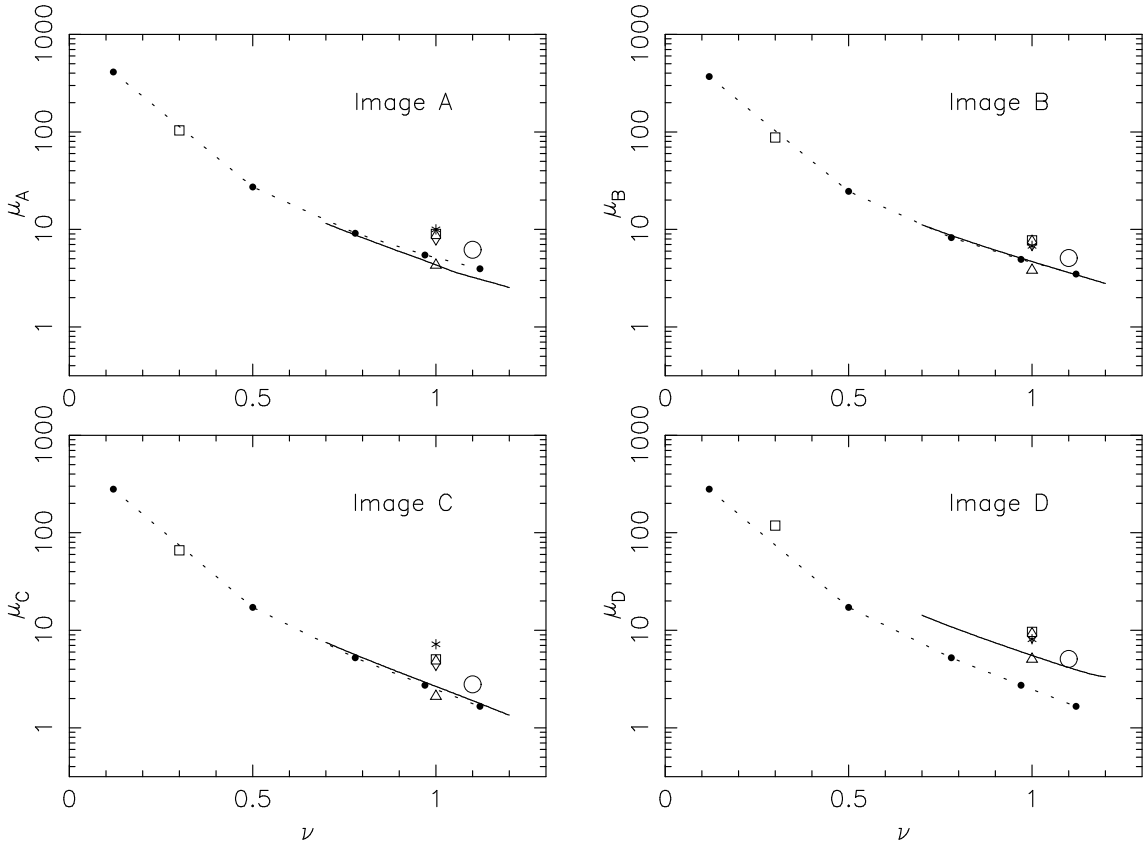}
\caption{The dependence of magnification $\mu$ on power-law index $\nu$. The circle represents model 2a from RSB92, the asterix, triangle and diamonds represent the models from K91 having internal, mixed and external shear respectively, and squares models from WP94. The dotted line joins points for models with zero core radius from CTK98, and the dark lines show values computed from the model of SWL98.}
\label{plot4} 
\end{figure*}

\begin{figure*}
\vspace*{135mm}
\includegraphics{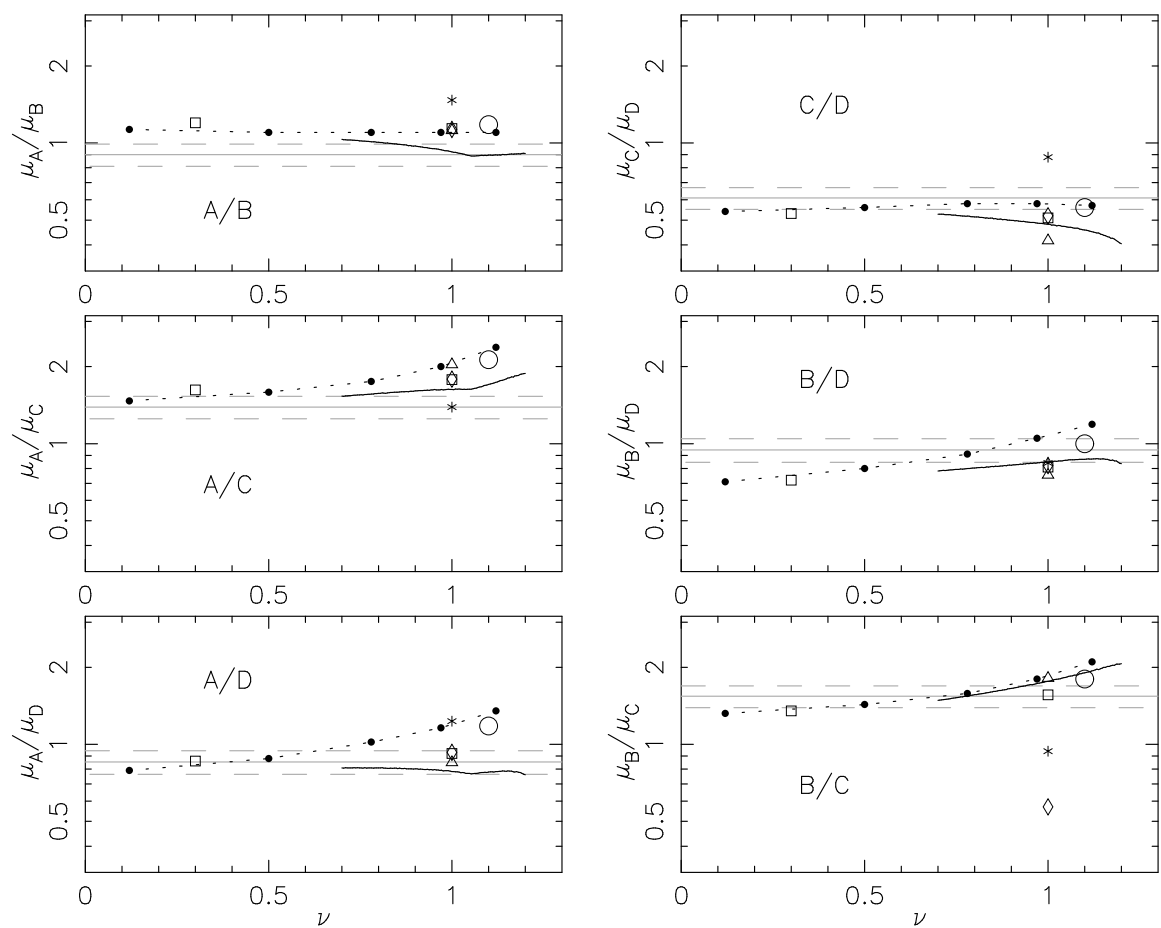}
\caption{The dependence of flux ratios $R$ on power-law index $\nu$. The circle represents model 2a from RSB92, the asterix, triangle and diamonds represent the models from K91 having internal, mixed and external shear respectively, and squares models from WP94. The dotted line joins points for models with zero core radius from CTK98, and the dark lines show values computed from the model of SWL98. The solid and dashed horizontal lines represent the measurement and $1\sigma$ uncertainty of the mid-IR flux ratios from November 2000 (Agol et al. 2001). }
\label{plot5} 
\end{figure*}

\begin{figure*}
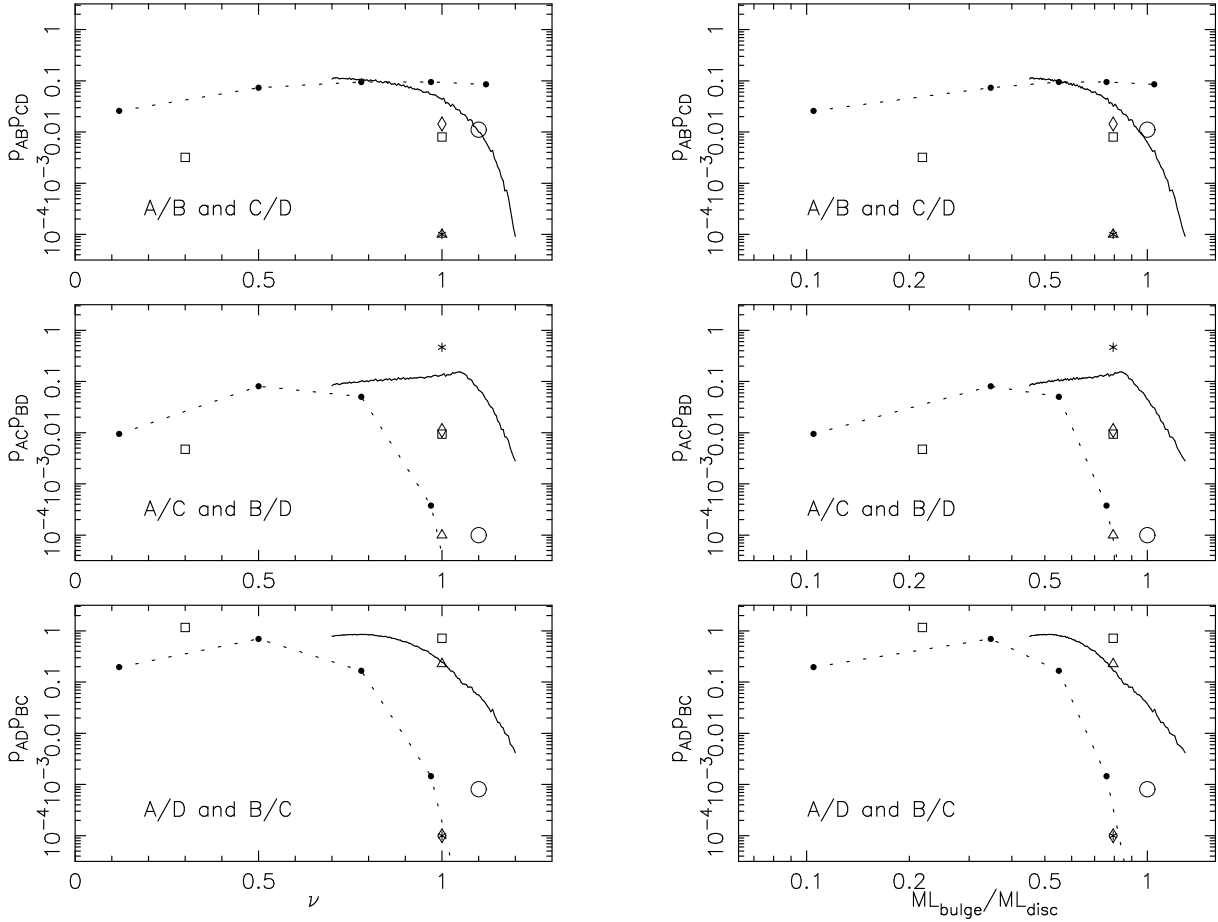

\vspace*{135mm}
\includegraphics{f6a.epsi}
\includegraphics{f6b.epsi}
\caption{Likelihoods for $\nu$ (left) and $ML_{bulge}/ML_{disc}$ (right). Two independent likelihoods have been combined in each case. Values of likeli-hood smaller than $10^{-4}$ have been plotted at the $10^{-4}$ level. The circle represents model 2a from RSB92, the asterix, triangle and diamonds represent the models from K91 having internal, mixed and external shear respectively, and squares models from WP94. The dotted line joins points for models with zero core radius from CTK98, and the dark lines show values computed from the model of SWL98. }
\label{plot6} 
\end{figure*}

Many authors have attempted to explain the geometry of Q2237+0305 via gravitational lensing using various models (summarised in SWL88 and WAF01). In this paper we consider the physically motivated parametric model of SWL98 which includes a bar component to correct for the apparent misalignment between the major axis of elliptical power-law models and the observed galactic orientation (e.g. K91; CTK98). SWL98 found a degeneracy between the ellipticity and slope of the mass profile (this degeneracy is a generic feature of parametric models (described by WP94)). For comparison, we also consider models from four other studies. Firstly, the model of RSB92 who assumed that mass in the lensing galaxy was proportional to observed light. We use their model 2a, which did not employ flux ratios as constraints. Secondly we consider the model of CTK98 who used a tri-axial power-law mass profile with a core (the position angle of this model is offset from the galaxy inclination axis by $\sim10^{\circ}$). CTK98 incorporated the 3.6cm fluxes measured by Falco et al. (1996). However the error-bars on these ratios were too large to break the degeneracy between ellipticity and $\nu$. In addition we consider models having a power-law mass-distribution with an external shear from WP94, and the isothermal models with internal, mixed and external shears from K91.

 Fig.~\ref{plot4} shows magnification as a function of projected power-law mass-function index $\nu$ for the models discussed. The circle represents the constant mass-to-light model of RSB92 (the value of $\nu$ was inferred from the photometric deconvolution of Schmidt 96). The asterix, triangle and diamonds represent the models from K91 having internal, mixed and external shear respectively, and squares models from WP94. The dark lines show results computed using the model described by SWL98, and the dotted lines join points for published results from CTK98 (for a zero core radius). The large dependence of the magnification $\mu$ on $\nu$ (WP94; SWL98; CTK98) is clearly evident from this figure. Six possible flux ratios have been calculated from the 4 images (Fig.~\ref{plot5}). The dependence of $R$ on $\nu$ is not as strong as that of $\mu$, however $R$ is directly observable.

Constraints on galaxy models for Q2237+0305 are obtained by comparing the predicted flux ratios with observations made in the mid-IR. Mid-IR flux ratios are usefull for two reasons. Firstly extinction, is not important in the mid-IR. Secondly, mid-IR emission is thought to come from an extended region, and should not be subject to microlensing (Agol, Jones \& Blaes 2000). Both these issues plauge broad-band optical determinations of the flux-ratios. The solid horizontal lines in  Fig.~\ref{plot5} are the observed mid-IR (November 2000) flux ratios (Agol et al. 2001). Since there was no pattern noise in these observations, the errors are assumed Gaussian (the horizontal dashed lines in Fig.~\ref{plot5} show the 1$\sigma$ errors).  There is a large range in predicted flux-ratio between different models, however Fig.~\ref{plot5} shows model flux ratios that differ less from the mid-IR observations for mass profiles which are shallower than isothermal. Likelihoods for $\nu$ were constructed for each image ratio, and combined for independent image pairs (Fig.~\ref{plot6}, left-hand panel). The flux ratios predicted by the models of CTK98 suggest that an isothermal profile is 3-4 orders of magnitude less likely than a mass profile with $\nu=0.5$. Less stringent limits are obtained from the models of SWL98. However, a value of $\nu=0.8$ is favoured over $\nu=1.2$ by more than a factor of $10^{2}-10^{3}$. Note also that the mid-IR flux ratios push the solution towards the deeper part of the $\chi^2$ valley shown in Fig. 3 of SWL98. The astigmatic models from WP94, the constant mass-to-light model from RSB92, and the isothermal models of K91 are strongly disfavoured with respect to the elliptical power-law models of SWL98 and CTK98.  

While the approach described does not constitute a self consistent inclusion of the flux ratios as model constraints, it is reasonable for two reasons. Firstly, the possible effects of microlensing are difficult to consistently include in a minimisation scheme (see the following section). Furthermore, since the relative errors in the IR fluxes are much larger than the errors in position used to constrain published models, consistent inclusion of the $R$s as constraints would leave the solution in the same $\chi^2$ valley in $\nu$-ellipticity space. 

Schmidt (1996) deconvolved HST photometry of the lens-galaxy into components of de~Vaucouleurs bulge, exponential disc, and Ferrers ellipse bar. At the location of the images, most light is contributed by the bulge and disc components, and we consider only them (results from Trott 2000 suggest that the halo component is also negligible at the location of the images). Furthermore, the location of the images coincide with the region where the contributions of the bulge and disc components are comparable. Assuming that the exponential disc profile can be interpolated into the inner kpc, the slope of the mass distribution can be re-cast as a ratio of mass-to-light between the bulge ($ML_{bulge}$) and disc ($ML_{disc}$) components. The right hand plot of Fig.~\ref{plot6} shows the limits re-displayed with $\nu$ replaced by $ML_{bulge}/ML_{disc}$. The mass-to-light ratio is likely to be higher in the disc than in the bulge.

\section{ The effect of microlensing}

\label{MLlim}

In this section we consider whether microlensing of the mid-IR emission region is an important factor for constraining the slope of the lens galaxy from mid-IR flux ratios. Sec.~\ref{IRlimits} discusses the probability of the size of the mid-IR emission region. In Sec.~\ref{MLconstraints} these source-size probabilities, are used to compute likeli-hoods for the flux ratios that include possible microlensing variability.

\subsection{Limits on the mid-IR source size from two epochs of data}
\label{IRlimits}

\begin{figure*}
\vspace*{130mm}
\includegraphics{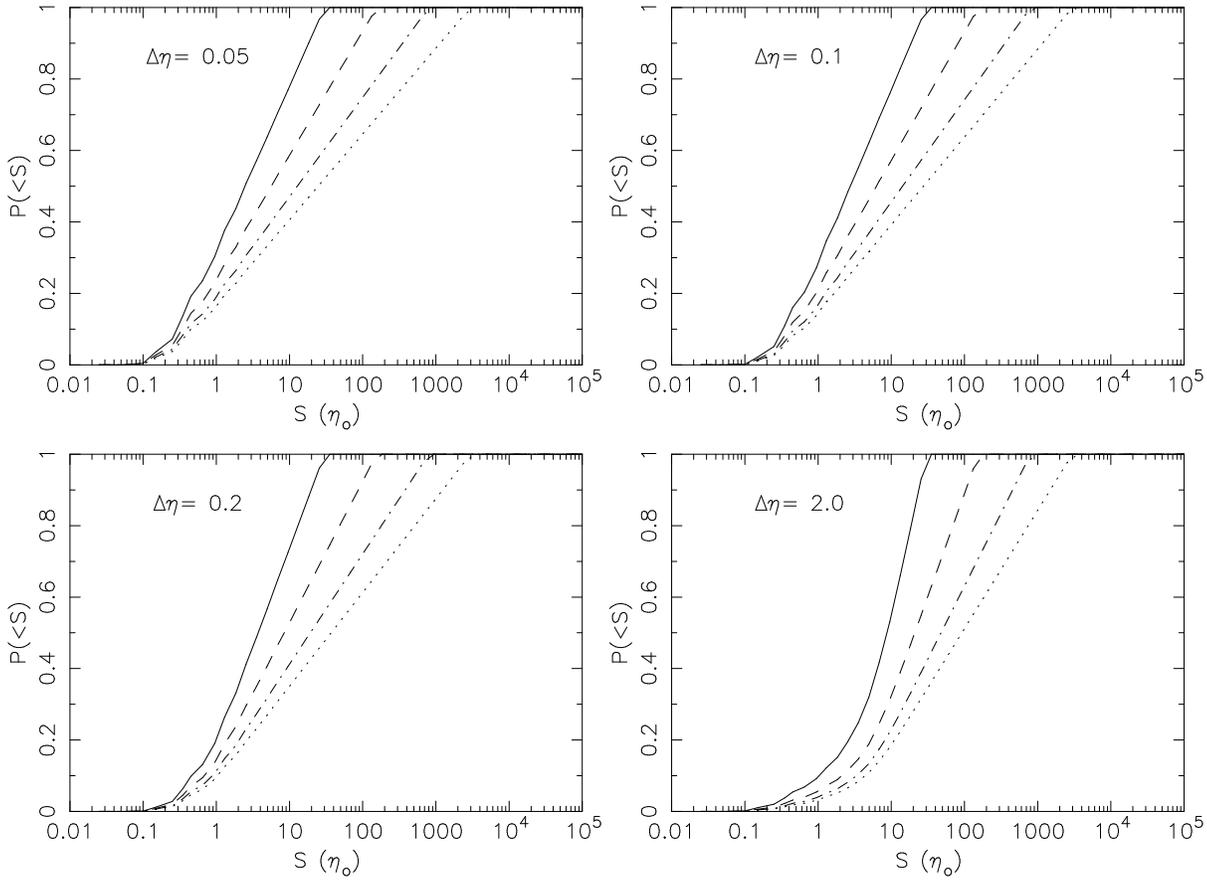}
\caption{The cumulative probability for the IR source size $P\left(S_{IR}<S|S_{OPT}\right)$ given a uniform optical source: $S_{OPT}= 0.025\eta_o$ (single pixel). The mid-IR source was Gaussian with halfwidth $\sigma=S_{IR}/2$, and the parameters for image C were $\kappa=0.69$, $\gamma=0.71$. Plots are shown assuming 4 cutoffs ($S_{max}$) on the logarithmic Bayesian prior for $S_{IR}$ (20$\eta_o$, solid lines; 100$\eta_o$, dashed lines; 500$\eta_o$, dot-dashed lines; 2500$\eta_o$, dotted lines). The 4 panels show limits assuming $\Delta \eta=0.05$, $\Delta \eta=0.10$, $\Delta \eta=0.20$ and $\Delta \eta=2.0$.}
\label{plot1} 
\end{figure*}

\begin{figure*}
\vspace*{65mm}
\includegraphics{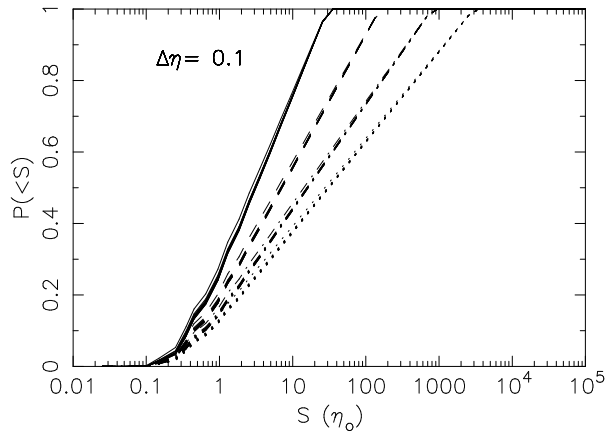}
\caption{The cumulative probability for the IR source size $P\left(S_{IR}<S|S_{OPT}\right)$ given a uniform optical source: $S_{OPT}= 0.025\eta_o$ (single pixel), and a Gaussian mid-IR source with halfwidth $\sigma=S_{IR}/2$. The macrolensing parameters were for models $i-xi$ described in WAF01. Plots are shown assuming 4 cutoffs ($S_{max}$) on the logarithmic Bayesian prior for $S_{IR}$ (20$\eta_o$, solid lines; 100$\eta_o$, dashed lines; 500$\eta_o$, dot-dashed lines; 2500$\eta_o$, dotted lines). The epoch separation was $\Delta \eta=0.10$.}
\label{plot3} 
\end{figure*}

Two epochs of mid-IR observations of Q2237+0305 have been made in September 1999 (AJB00) and November 2000 (Agol et al. 2001). Both epochs show significant variation in flux ratios between the mid-IR observations and the V-band observations described by Wozniak et al. (2000). The 1999 data have previously been used to place limits on the physical scale of mid-IR emission (AJB00; WAF01). Briefly, the argument is as follows. The optical source is known to undergo significant microlensing (Irwin et al. 1989; Corrigan et al. 1991; $\O$stensen et al. 1996; Wozniak et al. 2000a,b) and the variability record has been used to demonstrate that the optical source must be significantly smaller than the microlens Einstein Radius (e.g. Wambsganss, Paczynski \& Schneider 1990; Wyithe, Webster, Turner \& Mortlock 2000). Therefore at any one epoch, the optical flux ratios may differ significantly form the theoretical mean value (predicted from the galaxy mass model). If the mid-IR source is larger than the optical region, the mid-IR flux ratios will differ from those in the optical, and lie closer to the theoretical value. If the mid-IR source is much larger than the microlens Einstein Radius, then the mid-IR flux ratios will equal the theoretical value. In combination with an assumed macro-model, the flux-ratios can therefore be used to limit the size of the mid-IR emission region. The limits obtained using this method are sensitive to the average flux ratio and therefore to the macro-model assumed. However a similar argument applies to the flux ratios between observations of the same image at two different epochs, and this approach is pursued below. The flux levels of image C for both the September 1999 and November 2000 epochs are summarised in Tab.~\ref{tab1}. The Image C optical flux decreased significantly between September 1999 and November 2000, while the mid-IR flux remained steady. This temporal variability in the flux-ratio can be used to place limits on the mid-IR source size which are much less systematically dependent on the macro-model. Here we adapt the method described in detail in WAF01 for two different image ratios at the same epoch to calculate limits for the same image ratio at two different epochs. Because the epochs are separated by only 1 year, a combination of transverse velocity and microlens mass (expressed in units of the microlens Einstein Radius $\eta_o$) must be assumed. Note also that differential reddening of the V-band data need not be considered since the fluxes are measured from the same image.

We have combined likelihoods $\frac{dP_{lh}}{dR_{IR}}(R_{IR}|S',S_{OPT})$ for the mid-IR flux ratio $R_{IR}$ given the optical flux ratio $R_{OPT}$ (both measured for image C between 1999 and 2000), and mid-IR source size $S_{IR}$ (see WAF01) with uniform logarithmic Bayesian priors
\begin{eqnarray}
\label{prior}
\nonumber
\frac{dP_{prior}}{dlog(S_{IR})}&\propto& 1  \hspace{5mm} {\rmn where} \hspace{5mm} S_{IR}<S_{max}\\
                               &=& 0 \hspace{5mm} {\rmn otherwise}
\end{eqnarray}
for the unknown mid-IR source size $S_{IR}$:
\begin{eqnarray}
\label{size_lim}
\nonumber 
&&P_{S}(S<S_{IR}|S_{OPT}) = \\
&&N\int_0^{S_{IR}}dS'\frac{dP_{prior}}{dS}  \frac{dP_{lh}}{dR_{IR}}(R_{IR}|S',S_{OPT}).
\end{eqnarray}
$N$ is a normalising constant. Fig.~\ref{plot1} shows the resulting cumulative distributions $P_S(S_{IR}<S|S_{OPT})$ for $S_{IR}$ assuming the image C macro parameters calculated by SWL98 (model $xi$ from WAF01, $\kappa=0.69, \gamma=0.71$). A uniform optical source size was used: $S_{OPT}=0.025\eta_o$ (single pixel). The distributions are plotted assuming 4 different upper cutoffs of the Bayesian prior, $S_{max}=20, 100, 500$ and $2500\eta_o$. Four cases of epoch separation are shown $\Delta\eta_o=0.05$, $\Delta\eta_o=0.10$, $\Delta\eta_o=0.20$ and $\Delta\eta_o=2.0$. These correspond to galactic transverse velocities of $\sim$200, 400, 800 and 8000 $km/sec\left(\frac{\langle m\rangle}{m_{\odot}}\right)$, where $\langle m\rangle$ is the average microlens mass. The large separation corresponds to an unreasonably large velocity or small microlensing mass, and is included for comparison as it approximates the case of two random epochs. We interpret the 1999 image C peak as a cusp event (Wyithe, Webster \& Turner 2000). The $\sim 1$ year duration of this peak therefore indicates that a few tenths of an Einstein radius separate the 1999 and 2000 observations. Fig.~\ref{plot1} shows little variation in the limits obtained for values of separation in this range, and we therefore calculate limits at individual transverse velocities rather than introduce an additional prior.

Because both flux ratios are measured from the same image, the average flux ratio equals 1 for all models. The limits obtained should therefore be less dependent on macro-model than those presented in WAF01. Fig.~\ref{plot3} shows cumulative distributions assuming $\Delta\eta_o=0.1$,  $S_{OPT}=0.025\eta_o$ (single pixel) and the same 4 values of $S_{max}$. Distributions are shown for each of the macro-models $i-xi$ described in WAF01. These models have mean magnifications ranging from 2.72-5.56, and the lack of dependence of the distributions on the macro-model suggests that the mid-IR source size limits obtained apply to all reasonable macro-models for Q2237+0305.

 We find $S_{IR}>0.5\eta_o$ with greater than 90\% ($S_{OPT}=0.025\eta_o$) confidence. As a result of the smaller flux-ratio variation, this limit is not as strong as those presented in WAF01. Since the mid-IR did not vary by more than the observational uncertainty, the data carries no information upper limits, which depend on $S_{max}$. $S_{max}$ is the parameter to which the probabilities are most sensitive.

\subsection{ model constraints including microlensing}
\label{MLconstraints}

\begin{figure*}
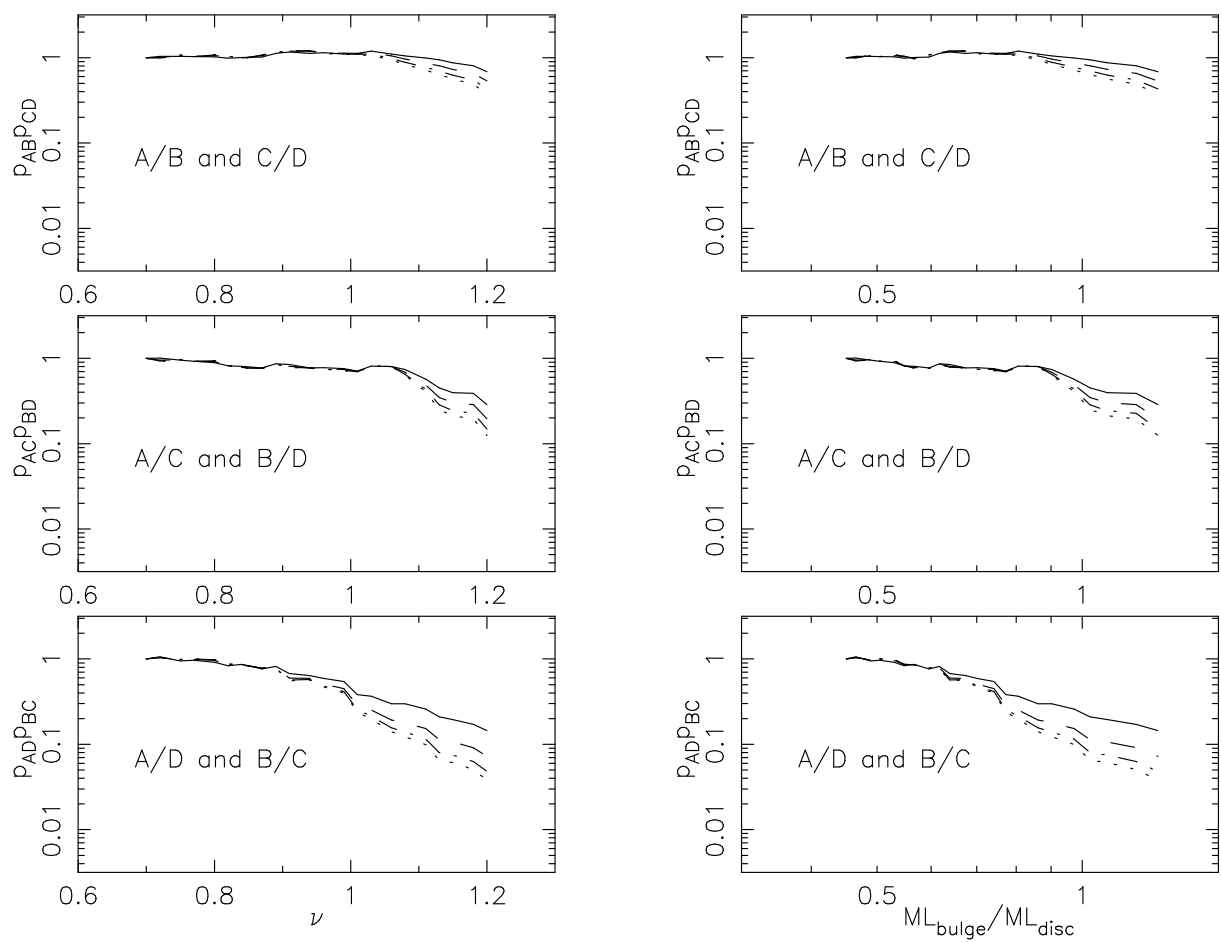

\vspace*{135mm}
\includegraphics{f7a.epsi}
\includegraphics{f7b.epsi}
\caption{Likelihoods for $\nu$ (left) and $ML_{bulge}/ML_{disc}$ (right) including the contribution of microlensing. Plots are shown assuming 4 cutoffs ($S_{max}$) on the logarithmic prior for $S_{IR}$ (20$\eta_o$, solid lines; 100$\eta_o$, dashed lines; 500$\eta_o$, dot-dashed lines; 2500$\eta_o$, dotted lines). Two independent likelihoods have been combined in each case. The macro models were computed according to SWL98. }
\label{plot8} 
\end{figure*}

\begin{figure*}
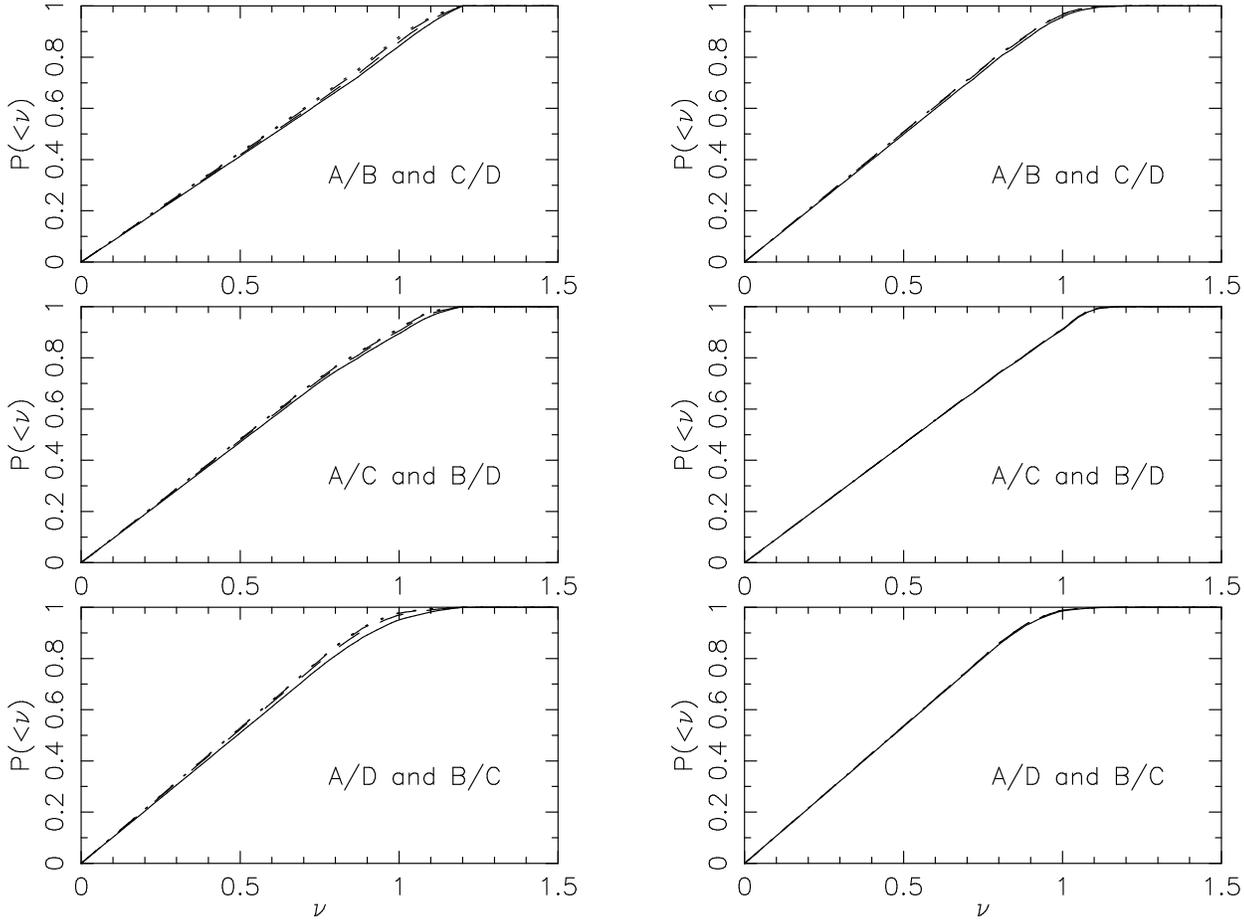

\vspace*{135mm}
\includegraphics{f8a.epsi}
\includegraphics{f8b.epsi}
\caption{A-posteriori probability for $\nu$ (including the contribution of microlensing). A flat Bayesian prior is assumed between $\nu=0$ and $\nu=1.3$. The likelihoods are assumed constant below and including $\nu=0.7$. Plots are shown assuming 4 cutoffs ($S_{max}$) on the logarithmic prior for $S_{IR}$ (20$\eta_o$, solid lines; 100$\eta_o$, dashed lines; 500$\eta_o$, dot-dashed lines; 2500$\eta_o$, dotted lines). Two independent likelihoods have been combined in each case. The macro model was described in SWL98. Two figures are shown. The left-hand plots are computed from the likeli-hoods in Fig.~\ref{plot8}. The right hand plots show the corresponding limits where the calculations in Figs.~\ref{plot1}-\ref{plot8} have been re-computed using a lower cut-off on the prior for mid-IR source size of 10$\eta_o$.}
\label{plot10} 
\end{figure*}

The results presented in Sec.~\ref{macros} assume that the mid-IR flux ratios accurately represent the true flux ratios (and hence that microlensing of the mid-IR source is negligible). However in Sec.~\ref{IRlimits}, we discussed macro-model insensitive limits on the mid-IR source size and it was shown that while this should be larger than $\eta_o$, there is some probability that the source is small enough to be affected by microlensing. In this section we consider how microlensing of the mid-IR affects the constraints imposed on the models of SWL98 by the mid-IR flux ratios. We have used a ray-tracing code generously provided by Joachim Wambsganss to compute magnification maps corresponding to the macro-parameters for each image at 22 different values of $\nu$. Three sets of magnification maps were computed to reduce numerical noise. Because the distributions of mid-IR source size shown in Fig.~\ref{plot1} are not sensitive to the macro-model, they may be used to compute a series of source size weighted flux-ratio distributions for each $\nu$. These distributions have been combined with the mid-IR observations to produce likelihoods for $\nu$ and $ML_{bulge}/ML_{disc}$ (Figs.~\ref{plot8}). Plots are shown assuming an optical source size $S_{OPT}=0.025\eta_o$ (single pixel), a separation of $\Delta \eta=0.1$ and the 4 Bayesian priors for $S_{IR}$ discussed in Sec.~\ref{IRlimits}. The inclusion of microlensing weakens the dependence of the likeli-hood on the slope of the mass-distribution. It may therefore be said that, in terms of the flux ratios the effect of microlensing of the mid-IR emission region is important, and this renders self consistent inclusion of the mid-IR flux ratios as model constraints difficult.  We note that the observed mid-IR flux ratios have been used both to estimate the source size, and using the probabilities obtained, to place limits on $\nu$. The two sets of limits are therefore not independent. However, the overlap is minimal since only the 1999 and 2000 image C mid-IR fluxes were used to obtain the source size limits, while all image fluxes from 2000 have been used to compute likelihoods for $\nu$.

The left-hand panel in Fig.~\ref{plot10} shows the a-posteriori probabilities for $\nu$ assuming the likelihoods in Fig.~\ref{plot8}, and a Bayesian prior that is flat between $\nu=0$ and $\nu=1.3$. The likelihood appears to flatten below $\nu\sim0.8-1.0$, and we assume it to be constant below and including $\nu=0.7$. The limits from the combination of image ratios A/D and B/C rule out a mass density that is steeper than isothermal with $\sim 90\%$ confidence. If the mid-IR emission is due to dust then it should be emitted from a region $>$ 1pc. We have recomputed the results in Figs.~\ref{plot1}-\ref{plot8} using a prior for mid-IR source size having a lower cut-off of 10$\eta_o$ ($\sim0.3\frac{\langle m\rangle}{1_{M_\odot}}$pc). The exclusion of the possibility for small sources limits the microlensing variability, and results in stronger limits of $\nu<0.8$ with 90\% confidence.

\section{Conclusion}

Mid-IR flux ratios (Agol et al. 2001) have been used to constrain degeneracies in the slope the mass profile for published models (Chae, Turnshek \& Khersonsky 1998; Kochanek 1991; Rix, Schneider \& Bahcall 1992; Schmidt, Webster \& Lewis 1998; Wambsganss \& Paczynski 1994) of the bulge of the lensing galaxy G2237+0305. Microlensing of the mid-IR is at a much lower level than for optical emission but may also be important. For the model of Schmidt, Webster \& Lewis 1998 our calculation includes microlensing of the mid-IR emission by first finding the a-posteriori probability for source size in a macro-model insensitive way using the variation of the image C V-band to mid-IR flux ratio between 2000 (Agol et al. 2001) and 1999 (Agol, Jones \& Blaes 2000). Microlensing induced flux ratio distributions were produced by combining distributions for the mid-IR source size with magnification distributions corresponding to macro-models having different power law slopes. Likelihoods for the slope of the mass profile were then computed from the combination of mid-IR flux ratio observations (with Gaussian errors) and microlensing induced flux ratio distributions. The mid-IR flux ratios favour a mass profile which is flatter than isothermal, resulting in an ellipticity lower than the light profile. Following the photometric disc-bulge deconvolution of Schmidt (1996), this can be interpreted as a mass-to-light ratio which is higher in the galactic disc than in the bulge. A shallow mass profile results in high magnification, and this helps alleviate the difficulty of reconciling the observed quasar luminosity with microlensing imposed source size constraints (Rauch \& Blandford 1991).

 Future mid-IR monitoring with reduced uncertainties will improve constraints on lens galaxy models for 2237+0305 in two ways. Firstly, if no variability of the mid-IR is observed, limits on the mid-IR source size will become tighter, and microlensing a less important consideration. On the other hand, if variability is observed there will no longer be strong dependence of mid-IR source size limits on the upper cutoff of Bayesian prior. Secondly, at the current level of uncertainty, the observations differ by only $\sim3\sigma$ from the predicted flux ratios for the least favoured models (from Schmidt, Webster \& Lewis 1998). Reduced uncertainties will therefore increase the disparity in the likelihoods of different models and result in tighter constraints

\section*{Acknowledgements}

The authors would like to thank Professor Joachim Wambsganss for providing his ray-tracing code. Some of the simulations presented in this paper were performed by Dr Chris Fluke at the Centre for Astrophysics and Supercomputing Swinburne University of Technology. This work was supported in part by NASA grant NAG5-9274 to ELT. JSBW acknowledges the support of an Australian Postgraduate Award.

\label{lastpage}

\end{document}